\documentclass[12pt]{article}
\newenvironment{eqnarraz}%
   {\setlength{\arraycolsep}{0.15em} \begin{eqnarray}}%
   {\end{eqnarray}}
\begin{document}
\begin{center}
{\LARGE \bf Lie Symmetries of the Self-Dual
Yang-Mills Equations}\\[5mm]
Marc Voyer and Louis Marchildon$^*$\\
D\'{e}partement de physique, Universit\'{e} du Qu\'{e}bec,\\
Trois-Rivi\`{e}res, Qu\'{e}bec, Canada G9A 5H7\\
$^*$e-mail: marchild@uqtr.uquebec.ca\\[5mm]
\end{center}
\begin{abstract}
We investigate Lie symmetries of the self-dual
Yang-Mills equations in four-dimen\-sional
Euclidean space (SDYM).
The first prolongation of the symmetry generating vector fields
is written down, and its action on SDYM computed.  Determining
equations are then obtained and solved completely.  Lie
symmetries of SDYM in Euclidean space are in exact correspondence
with symmetries of the full Yang-Mills equations in Minkowski
space.
\end{abstract}
\section{Introduction}
The self-dual Yang-Mills equations in Euclidean space (SDYM) are
the following first-order nonlinear partial differential
equations:
\begin{equation}
\left( \delta_{\mu \rho} \delta_{\nu \sigma} - \frac{1}{2}
\epsilon_{\mu \nu \rho \sigma} \right) \left( \partial_{\rho}
A_{a \sigma} - \partial_{\sigma} A_{a \rho} + C_{abc} A_{b \rho}
A_{c \sigma} \right) = 0 . \label{sdym}
\end{equation}
Greek indices label independent variables $x_{\nu}$ and run from
0 to 3.  Latin indices are associated with generators of a
compact semisimple Lie algebra, with structure constants
$C_{abc}$.  $\delta_{\mu \rho}$ and $\epsilon_{\mu \nu \rho
\sigma}$ are the Kronecker and Levi-Civita symbols
($\epsilon_{0123} = 1$), respectively.  The $A_{a \sigma}$ are
dependent variables, namely, gauge potentials.

We shall be interested in Lie symmetries of Eq.~(\ref{sdym}).
Consider a vector field $v$ of the form
\begin{equation}
v = H_{\kappa} \partial_{\kappa} + \Phi_{d \kappa}
\frac{\partial}{\partial A_{d \kappa}} , \label{vector}
\end{equation}
where $H_{\kappa}$ and $\Phi_{d \kappa}$ are functions of
$x_{\nu}$ and $A_{a \sigma}$.  The first prolongation of $v$ is
defined as
\begin{equation}
\mbox{pr} ^{(1)} v = H_{\kappa} \partial_{\kappa} + \Phi_{d
\kappa} \frac{\partial}{\partial A_{d \kappa}} + \Phi_{d \kappa
\lambda} \frac{\partial}{\partial (\partial_{\lambda} A_{d
\kappa})} , \label{prv}
\end{equation}
where $\Phi_{d\kappa \lambda}$ is given by
\begin{equation}
\Phi_{d  \kappa \lambda} = \partial_{\lambda} \Phi_{d  \kappa} -
(\partial_{\lambda} H_{\beta}) \partial_{\beta} A_{d \kappa} +
(\partial_{\lambda} A_{n \alpha}) \frac{\partial \Phi_{d
\kappa}}{\partial A_{n \alpha}} - (\partial_{\lambda} A_{n
\alpha}) (\partial_{\beta} A_{d \kappa}) \frac{\partial H_{
\beta}}{\partial A_{n \alpha}} . \label{phi}
\end{equation}
To obtain Lie symmetries of SDYM, we have to substitute
(\ref{phi}) in (\ref{prv}) and let it act on (\ref{sdym}).  We
obtain
\begin{eqnarray}
\lefteqn{\left( \partial_{\lambda} \Phi_{a \kappa} + C_{abc} A_{b
\lambda} \Phi_{c \kappa} \right) Z_{\mu \lambda \nu \kappa} -
(\partial_{\lambda} A_{n \alpha}) (\partial_{\beta} A_{a
\kappa}) \frac{\partial H_{\beta}}{\partial A_{n \alpha}} Z_{\mu
\lambda \nu \kappa}} \nonumber \\
& & \mbox{} + (\partial_{\lambda} A_{n \alpha}) \left[
\frac{\partial \Phi_{a \kappa}}{\partial A_{n \alpha}} Z_{\mu
\lambda \nu \kappa} - (\partial_{\kappa}
H_{\lambda}) Z_{\mu \kappa
\nu \alpha} \delta_{an} \right] = 0 , \hspace{1cm}
\label{action}
\end{eqnarray}
where
\begin{equation}
Z_{\mu \lambda \nu \kappa} = \delta_{\mu \lambda} \delta_{\nu
\kappa} - \delta_{\mu \kappa} \delta_{\nu \lambda}  -
\epsilon_{\mu \nu \lambda \kappa} .
\end{equation}
The vector field $v$ generates a symmetry of SDYM provided that
Eqs.~(\ref{action}) hold whenever SDYM hold \cite{olver}.  In
other words, once SDYM are substituted in (\ref{action}), the
coefficients of each independent combination of derivatives of
$A_{a \sigma}$ must vanish.  Note that SDYM can be written more
explicitly as
{\begin{eqnarraz}
\partial_2 A_{n3} - \partial_3 A_{n2} &=& \partial_0 A_{n1} -
\partial_1 A_{n0} + C_{nbc} (A_{b0} A_{c1} + A_{b3} A_{c2} ) ,
\label{sdymx1} \\
\partial_3 A_{n1} - \partial_1 A_{n3} &=& \partial_0 A_{n2} -
\partial_2 A_{n0} + C_{nbc} (A_{b0} A_{c2} + A_{b1} A_{c3} ) ,
\label{sdymx2} \\
\partial_1 A_{n2} - \partial_2 A_{n1} &=& \partial_0 A_{n3} -
\partial_3 A_{n0} + C_{nbc} (A_{b0} A_{c3} + A_{b2} A_{c1} ) .
\label{sdymx3}
\end{eqnarraz}}%
\section{Determining equations}
SDYM only involve combinations of derivatives
$\partial_{\lambda} A_{n \alpha}$ that are antisymmetric in
$\lambda$ and $\alpha$.  Coefficients of symmetric combinations
must therefore vanish.  Accordingly, we set to zero the
coefficient of $(\partial_{\lambda} A_{n \alpha})
(\partial_{\beta} A_{m \kappa})$, symmetrized in $\lambda
\leftrightarrow \alpha$, in $\beta \leftrightarrow \kappa$ and,
furthermore, in $(\lambda n \alpha) \leftrightarrow (\beta  m
\kappa)$. The result is, $\forall \mu$, $\nu$, $\lambda$,
$\kappa$, $\beta$, $\alpha$, $a$, $m$ and $n$
\begin{eqnarray}
\lefteqn{\delta_{am} \left\{ \frac{\partial H_{\beta}}{\partial
A_{n \alpha}} Z_{\mu \lambda \nu \kappa} + \frac{\partial
H_{\beta}}{\partial A_{n \lambda}} Z_{\mu \alpha \nu \kappa} +
\frac{\partial H_{\kappa}}{\partial A_{n \alpha}} Z_{\mu \lambda
\nu \beta} + \frac{\partial H_{\kappa}}{\partial A_{n \lambda}}
Z_{\mu \alpha \nu \beta} \right\} } \nonumber \\
& & \mbox{} + \delta_{an}
\left\{ \frac{\partial H_{\lambda}}{\partial A_{m
\kappa}} Z_{\mu \beta \nu \alpha} + \frac{\partial
H_{\lambda}}{\partial A_{m \beta}} Z_{\mu \kappa \nu \alpha} +
\frac{\partial H_{\alpha}}{\partial A_{m \kappa}} Z_{\mu \beta
\nu \lambda} + \frac{\partial H_{\alpha}}{\partial A_{m \beta}}
Z_{\mu \kappa \nu \lambda} \right\} = 0 .
\label{dada1}
\end{eqnarray}
Taking $\mu$, $\nu$, $\lambda$ and $\kappa$ all different,
$\alpha = \lambda$ and $a = m \neq n$, we get
\begin{equation}
\epsilon_{\mu \nu \hat{\lambda}  \kappa} \frac{\partial
H_{\beta}}{\partial A_{n \hat{\lambda}}} + \epsilon_{\mu \nu
\hat{\lambda}  \beta} \frac{\partial H_{\kappa}}{\partial A_{n
\hat{\lambda}}} =  0 . \label{dada2}
\end{equation}
The hat means that the summation convention is not to be carried
over $\lambda$.  The last equation holds identically if and only
if, $\forall \beta$, $\lambda$ and $n$
\begin{equation}
\frac{\partial H_{\beta}}{\partial A_{n \lambda}} = 0 .
\label{det1}
\end{equation}
Thus in Eqs.~(\ref{action}), all terms quadratic in partial
derivatives of $A_{d \kappa}$ vanish.

Let us now switch to terms linear in partial derivatives.  The
coefficient of $\partial_{\lambda} A_{n \alpha}$, symmetrized in
$\lambda \leftrightarrow  \alpha$, must vanish.  Writing down the
coefficient, and considering in turn all independent values of
indices, one finds that conditions for this are that $\forall a$,
$n$ and $\forall \hat{\lambda}, \hat{\alpha}$
\begin{equation}
\frac{\partial \Phi_{a  \hat{\alpha}}}{\partial A_{n
\hat{\alpha}}} - \frac{\partial \Phi_{a  \hat{\lambda}}}{\partial
A_{n \hat{\lambda}}} + (\partial_{\hat{\alpha}} H_{\hat{\alpha}}
- \partial_{\hat{\lambda}} H_{\hat{\lambda}}) \delta_{an} = 0
\label{det2}
\end{equation}
and that, $\forall a$, $n$ and $\forall \nu, \alpha \neq$
\begin{equation}
\frac{\partial \Phi_{a \nu}}{\partial A_{n \alpha}} +
\partial_{\nu} H_{\alpha} \delta_{an} = 0 . \label{det3}
\end{equation}

For given $n$, of the six antisymmetric combinations
$\partial_{\lambda} A_{n \alpha} - \partial_{\alpha}  A_{n
\lambda}$, only three are independent, others being
constrained by Eqs.~(\ref{sdymx1})--(\ref{sdymx3}).  The choice
of independent combinations is arbitrary.  We pick $(\lambda,
\alpha) = (0,1)$, $(0,2)$ and $(0,3)$, and substitute dependent
combinations as given by (\ref{sdymx1})--(\ref{sdymx3}) in
(\ref{action}).  The coefficient of independent combinations must
then vanish.  After some calculation we find that, $\forall a$,
$n$ and $\forall \hat{\lambda}$, $\hat{\alpha}$
\begin{equation}
\frac{\partial \Phi_{a  \hat{\alpha}}}{\partial A_{n
\hat{\alpha}}} - \frac{\partial \Phi_{a  \hat{\lambda}}}{\partial
A_{n \hat{\lambda}}} - (\partial_{\hat{\alpha}} H_{\hat{\alpha}}
- \partial_{\hat{\lambda}} H_{\hat{\lambda}}) \delta_{an} = 0
\label{det4}
\end{equation}
and that, $\forall a$, $n$ and $\forall \lambda, \alpha \neq$
\begin{equation}
\frac{\partial \Phi_{a \alpha}}{\partial A_{n \lambda}} +
\frac{\partial \Phi_{a \lambda}}{\partial A_{n \alpha}} -
(\partial_{\alpha} H_{\lambda} + \partial_{\lambda} H_{\alpha})
\delta_{an} = 0 . \label{det5}
\end{equation}

There remain terms with no derivatives in $A_{d \kappa}$. To
terms in (\ref{action}) we must add the ones coming from the
substitution of SDYM just before Eq.~(\ref{det4}).  The sum must
vanish, yielding (indices $i$ and $j$ are summed from 1 to 3)
\begin{eqnarray}
\lefteqn{\left( \partial_{\lambda} \Phi_{a \kappa} + C_{abc} A_{b
\lambda} \Phi_{c \kappa} \right) Z_{\mu \lambda \nu \kappa} +
\frac{1}{4} C_{nbc} \left\{ A_{bi} A_{cj} - \frac{1}{2}
\epsilon_{i j \rho \sigma} A_{b \rho} A_{c \sigma}
\right\}} \nonumber \\
& &  \mbox{} \cdot \left\{ \left( \frac{\partial \Phi_{a
\kappa}}{\partial A_{n i}} - \partial_{\kappa} H_i \delta_{an}
\right) Z_{\mu j \nu \kappa} - \left( \frac{\partial \Phi_{a
\kappa}}{\partial A_{n j}} - \partial_{\kappa} H_j \delta_{an}
\right) Z_{\mu i \nu \kappa}
\right\} = 0 . \label{det6}
\end{eqnarray}
\section{Solution of Determining Equations}
We proceed to solve Eqs.~(\ref{det1}), (\ref{det2}),
(\ref{det3}), (\ref{det4}), (\ref{det5}) and (\ref{det6}).

Eqs.~(\ref{det1}) imply that $H_{\beta}$ is independent of $A_{n
\lambda}$, that is, $H_{\beta} = H_{\beta} (x_{\nu})$.  Combining
(\ref{det2}) with (\ref{det4}), we see that
$\partial_{\hat{\alpha}} H_{\hat{\alpha}}$ is independent of
$\hat{\alpha}$.  Combining (\ref{det3}) with (\ref{det5}), we
find that $\forall \lambda, \alpha \neq$, $\partial_{\alpha}
H_{\lambda} =  - \partial_{\lambda} H_{\alpha}$.  This means that
\begin{equation}
\partial_{\kappa} H_{\lambda} = f_{\lambda \kappa} +
\delta_{\lambda \kappa} G , \label{s1}
\end{equation}
where $G$ and $f_{\lambda \kappa} = - f_{\kappa  \lambda}$ are
arbitrary functions of $x_{\nu}$.  Coefficients of $\delta_{an}$
in (\ref{det2}), (\ref{det4}) and (\ref{det5}) have by now all
vanished.

From (\ref{det3}) we see that $\Phi_{a \kappa}$ is independent of
$A_{n \alpha}$ for $a \neq n$ and $\kappa \neq \alpha$.
Moreover, $\Phi_{a \kappa}$ is linear in $A_{a \alpha}$. With
(\ref{s1}), we can thus write
\begin{equation}
\Phi_{a \kappa} = f_{\kappa \alpha} (x_{\nu}) A_{a \alpha} + F_{a
\kappa} (A_{m \kappa}, x_{\nu}) . \label{s2}
\end{equation}
From (\ref{det4}), we see that $\forall a$ and  $n$,
$\partial \Phi_{a \hat{\alpha}} / \partial A_{n \hat{\alpha}}$ is
independent of $\hat{\alpha}$.  A little thought shows that
$\Phi_{a \kappa}$ must be linear in $A_{n \kappa}$, so that
\begin{equation}
\Phi_{a \kappa} = f_{\kappa \alpha} (x_{\nu}) A_{a \alpha} +
h_{an} (x_{\nu}) A_{n \kappa} + F_{a \kappa} (x_{\nu}) .
\label{s3}
\end{equation}
Eqs.~(\ref{s1}) and (\ref{s3}) are the most general solutions of
Eqs.~(\ref{det1}), (\ref{det2}), (\ref{det3}), (\ref{det4}) and
(\ref{det5}).

There remains to substitute (\ref{s1}) and (\ref{s3}) in
(\ref{det6}), which must hold as an identity.  In other words,
the coefficients of each combinations of $A_{d  \kappa}$ must
vanish.  After some manipulations, there result the following
equations:
\begin{eqnarray}
\partial_{\mu} F_{a \nu} - \partial_{\nu} F_{a \mu} -
\epsilon_{\mu \nu \lambda \kappa} \partial_{\lambda} F_{a \kappa}
= 0 & & \forall \mu, \nu, a; \label{e1} \\
\partial_{\mu} h_{\hat{a} \hat{a}} - \partial_{\hat{\nu}} f_{\mu
\hat{\nu}} -  \epsilon_{\mu \hat{\nu}  \lambda \alpha}
\partial_{\lambda} f_{\alpha \hat{\nu}} = 0 & & \forall \hat{a},
\forall \mu, \hat{\nu} \neq; \label{e2} \\
\partial_{\mu} h_{an} - C_{anc} F_{c \mu} = 0  & & \forall \mu,
\forall a, n \neq; \label{e3}\\
C_{abn} h_{nc} - C_{acn} h_{nb} - C_{nbc} h_{an} + C_{abc} G = 0
& & \forall a, b, c. \label{e4}
\end{eqnarray}

It was shown in \cite{marchildon} that the most general
solution of Eqs.~(\ref{e4}) is
\begin{equation}
h_{an} = - G \delta_{an} +  C_{anc}  \chi_c , \label{s4}
\end{equation}
where $\chi_c$ is an arbitrary function of $x_{\nu}$.
Substituting (\ref{s4}) in (\ref{e3}), we see that
\begin{equation}
F_{c \mu} = \partial_{\mu} \chi_c . \label{s5}
\end{equation}
Thus, (\ref{e1}) holds identically.  Substituting (\ref{s4}) in
(\ref{e2}) yields $\forall \mu, \hat{\nu} \neq$
\begin{equation}
- \partial_{\mu} G = \partial_{\hat{\nu}} f_{\mu \hat{\nu}} +
\epsilon_{\mu \hat{\nu} \lambda \alpha} \partial_{\lambda}
f_{\alpha \hat{\nu}} . \label{s6}
\end{equation}

There only remains to solve Eqs.~(\ref{s6}).  Note that they do
not involve group indices.  The general solution of (\ref{s6})
can be effected as follows.
\begin{enumerate}
\item Eliminate $G$ from the 12 Eqs.~(\ref{s6}) to obtain 8
equations for $p = f_{10} + f_{23}$, $q = f_{20} +  f_{31}$ and
$r  =  f_{30} + f_{12}$.  Show that all third derivatives of $p$,
$q$ and $r$ vanish and that (all constants being arbitrary)
{\begin{eqnarraz}
p &=& e + e_0 t + e_1 x + e_2 y + e_3 z + e_{00} (t^2 + x^2 - y^2
- z^2) \nonumber \\
& & \;\;\;\; \mbox{} + 2 e_{12} (xy + tz) + 2 e_{13} (xz - ty),
\label{e5} \\
q &=& e' + e_3 t - e_2 x + e_1 y - e_0 z + e_{12} (t^2 - x^2 +
y^2 - z^2) \nonumber \\
& & \;\;\;\; \mbox{} + 2 e_{13} (tx + yz) + 2 e_{00} (xy - tz),
\label{e6} \\
r &=& e'' - e_2 t - e_3 x + e_0 y + e_1 z + e_{13} (t^2 - x^2 -
y^2 + z^2) \nonumber \\
& & \;\;\;\; \mbox{} + 2 e_{12} (yz - tx) + 2 e_{00} (xz + ty).
\label{e7}
\end{eqnarraz}}%
\item With the help of Eq.~(\ref{s1}), express the 4 remaining
equations in terms of the functions $p$, $q$ and $r$ and of
second derivatives of $H_{\lambda}$.
\item Show that the resulting system has no solution for
$e_{00}$, $e_{12}$ or $e_{13}$ different from zero, and that its
most general solution is (with $b_{\lambda \alpha}$ antisymmetric
and all constants arbitrary)
\begin{equation}
H_{\lambda} =  - \frac{1}{2} c_{\lambda} x_{\alpha} x_{\alpha} +
c_{\alpha} x_{\lambda} x_{\alpha} + b_{\lambda \alpha} x_{\alpha}
+ d x_{\lambda} + a_{\lambda} . \label{s7}
\end{equation}
\end{enumerate}

Putting everything together, we find that
\begin{equation}
\Phi_{a \kappa} = (- c_{\kappa} x_{\alpha} +  c_{\alpha}
x_{\kappa} + b_{\kappa \alpha}) A_{a \alpha} - (d + c_{\alpha}
x_{\alpha}) A_{a \kappa} + C_{abd} \chi_d A_{b \kappa} +
\partial_{\kappa} \chi_a . \label{s8}
\end{equation}
Eqs.~(\ref{s7}) and (\ref{s8}) are the most general solution of
the determining equations.  Therefore, the corresponding vector
field (\ref{vector}) generates Lie symmetries of SDYM.  One can
see that the constants $a_{\mu}$ correspond to uniform
translations; that the $b_{\mu \nu}$ correspond to rotations in
Euclidean space; that the $c_{\mu}$ correspond to uniform
accelerations; that $d$ corresponds to dilatations; and that the
functions $\chi_a (x_{\nu})$ correspond to local gauge
transformations.  This agrees with results found in the special
case where the gauge group is SU(2)
\cite{rosenhaus,kersten}.  Therefore,
Lie symmetries of SDYM in Euclidean space correspond to
symmetries of the Yang-Mills equations in Minkowski space, which
are conformal transformations and gauge transformations
\cite{marchildon,torre}.
\end{document}